\documentclass{article}
\usepackage{spconf,amsmath,graphicx}
\usepackage{subcaption}
\usepackage{amssymb}
\usepackage[table]{xcolor}
\usepackage[nohyperlinks]{acronym}
\usepackage[disable,colorinlistoftodos]{todonotes}
\usepackage{bm}
\usepackage{booktabs}
\usepackage{multirow}
\usepackage{algorithm}
\usepackage{algorithmic}
\usepackage{tikz}
\usetikzlibrary{arrows.meta,positioning}
\definecolor{LinkBlue}{RGB}{0,70,160}
\definecolor{CiteBlue}{RGB}{0,90,170}
\definecolor{UrlBlue}{RGB}{0,90,200}
\definecolor{DarkGreen}{RGB}{34,139,34}

\usepackage[colorlinks=true, linkcolor=LinkBlue, citecolor=CiteBlue, urlcolor=UrlBlue]{hyperref}
\usepackage{cleveref}
\usepackage{soul}
\setuldepth{foobar}

\crefname{algorithm}{Alg.}{Algs.}
\crefname{equation}{Eq.}{Eqs.}
\crefname{figure}{Fig.}{Figs.}
\crefname{table}{Table}{Tables}
\crefname{section}{Section}{Sections}
\crefname{subsection}{Section}{Sections}

\def\modelName{SSE}
\def\modelNameLong{\emph{Spot, Separate, and Enhance}}
\acrodef{VAE}{Variational Autoencoder}
\acrodef{DL}{Deep Learning}
\acrodef{SSL}{Self-Supervised Learning}
\acrodef{\modelName}{\modelNameLong}
\acrodef{UGC}{User Generated Content}
\acrodef{RVQ}{Residual Vector Quantization}
\acrodef{CFM}{Conditional Flow Matching}
\acrodef{PE}{Perception Encoder}
\acrodef{DiT}{Diffusion Transformer}
\acrodef{KL}{Kullback-Leibler Divergence}
\acrodef{FD}{Fréchet Audio Distance}
\acrodef{IS}{Inception Score}
\acrodef{IB}{ImageBind Score}
\acrodef{Sync}{Audio-Visual Synchronization Score}
\acrodef{CLAP}{Semantical Language-Audio Alignment}
\acrodef{LoRA}{Low-Rank Adaptation}
\acrodef{MSE}{Mean Squared Error}
\acrodef{VGAH}{Visually Guided Acoustic Highlighting}

\usepackage{xspace}

\def\eqref#1{equation~\ref{#1}}
\def\1{\bm{1}}

\def\vh{{\bm{h}}}

\def\vv{{\bm{v}}}

\def\vx{{\bm{x}}}

\def\vz{{\bm{z}}}

\def\mC{{\bm{C}}}

\def\mI{{\bm{I}}}

\DeclareMathAlphabet{\mathsfit}{\encodingdefault}{\sfdefault}{m}{sl}
\SetMathAlphabet{\mathsfit}{bold}{\encodingdefault}{\sfdefault}{bx}{n}

\newcommand{\R}{\mathbb{R}}

\newcommand{\PAR}[1]{\vskip4pt
\noindent
{\bf #1~}}

\renewcommand{\paragraph}[1]{\vspace{.5em}
\noindent
\textbf{#1.}}

\usepackage{microtype}

\title{Spot, Separate, and Enhance: Fully Generative Approach for Audio Mixing}

\name{Ilpo Viertola$^{1,2}$\thanks{This work was conducted during an internship at Dolby Laboratories.}, Giulio Cengarle$^{1}$, Gouthaman KV$^{1}$, Daniel Arteaga$^{1}$, Lie Lu$^{1}$}
\address{$^{1}$Dolby Laboratories $^{2}$Tampere University}

\begin{document}
\ninept
\maketitle
\begin{abstract}
We introduce \ac{\modelName}, the first multimodal, user-guided generative model for audio remixing and enhancement.
\ac{\modelName} enhances video content by rebalancing the audio, removing unwanted audio sources, and reducing reverberation, guided by both video and textual descriptions.
To support its training and evaluation, we propose \emph{DegradedMix}, a new dataset built on the audio remixing benchmark MuddyMix.
We also adopt evaluation metrics from generative modeling, which better capture the creative nature of remixing than standard reconstruction-based metrics.
\ac{\modelName} outperforms existing baselines in both controllability and remixing quality, as shown by extensive experiments.
Project page: \href{https://sse-ai.notion.site/}{sse-ai.notion.site}
\end{abstract}
\begin{keywords}
Generative Learning, Audio Remixing, Multimodal Learning
\end{keywords}
\section{Introduction}
\label{sec:intro}
%%% ---- Section Content: Introduction ----
Recent advances in deep learning and generative modeling have opened new possibilities for enhancing \acf{UGC}.
A large portion of \ac{UGC} consists of videos captured in everyday environments, where audio often contains multiple sound sources, background noise, reverberation, and other recording artifacts.
Despite improvements in video capture hardware, post-processing remains essential for improving audio quality and intelligibility.
For example, rebalancing the audio mix to emphasize the primary subject, suppressing background noise, or mitigating reverberation can substantially improve the viewing experience.

To enable automatic audio remixing, Huang et al.~\cite{huang2025learning} introduced \acf{VGAH}, which rebalances the audio mix to emphasize the main subject based on visual information.
While an important step toward automated audio enhancement, \ac{VGAH} primarily considers loudness adjustments and does not address other common degradations such as background noise or reverberation.
Together with the task, Huang et al.~\cite{huang2025learning} proposed an evaluation framework based primarily on differences between the generated remix and ground-truth balanced mix in waveform or spectral representations.
However, audio remixing is inherently creative, with multiple perceptually valid solutions, and reconstruction-based evaluation may therefore penalize valid outputs that differ from a particular ground truth.
These limitations motivate a more general remixing framework that can address diverse degradations and acoustic conditions while allowing greater flexibility in the solution space.

Existing approaches \cite{huang2025learning,malard2026conditional} use an encoder-decoder model \cite{rouard2022hybrid} where a Transformer \cite{vaswani2017attention} processes the latent audio representation conditioned on video frames, and the decoder generates the remixed audio.
Huang et al. \cite{huang2025learning} propose VisAH, which is trained end-to-end in a discriminative manner using a reconstruction loss between the automatic remix and the ground truth balanced mix.
Malard et al. \cite{malard2026conditional} propose VisAH-FM, a concurrent work that adopts the same architecture with VisAH but replaces the discriminative training objective with a generative \ac{CFM} \cite{lipman2022flow} formulation.
They further introduce a rollout loss \cite{song2023consistency,song2024improved}, which performs full generation during training and computes an \acf{MSE} against the ground-truth target, exposing the model to its own intermediate predictions and encouraging self-correction.

However, both approaches have limitations.
First, VisAH inherently struggles to generate new audio as it predicts a spectral mask applied to the input mixture.
When the target source is heavily buried in noise or reverberation, the model cannot generate new content to replace the degraded audio.
Second, both approaches train the audio codec jointly with the generative backbone which limits scalability \cite{liu2024audioldm, cheng2025mmaudio}.
Third, neither approach provides user control beyond the video conditioning.
Finally, the rollout loss in VisAH-FM introduces overhead by requiring full generation for every training sample.

To address these limitations, we propose \acf{\modelName}, a multimodal, user-guided generative framework for audio remixing.
Unlike existing approaches that primarily modify the input mixture, \ac{\modelName} can generate new audio to replace severely degraded target-source regions, while explicit user guidance enables control over the desired remixing behavior.

We also introduce a new evaluation framework based on principles from generative modeling, which are better suited to a creative task such as audio remixing.
Rather than requiring close reconstruction of a single ground-truth target, our evaluation considers the perceptual quality and effectiveness of the resulting remix.
We further define a controlled set of audio degradations to simulate common real-world recording defects, enabling systematic evaluation under challenging acoustic conditions.
Extensive experiments demonstrate that \ac{\modelName} achieves superior remixing quality compared to existing approaches while providing greater user controllability and the ability to recover target-source content under severe audio degradation.

%%% ----------------------------------------

\section{Method}
\label{sec:method}
%%% ---- Section Content: Method ----
\begin{figure*}
    \centering
    \includegraphics{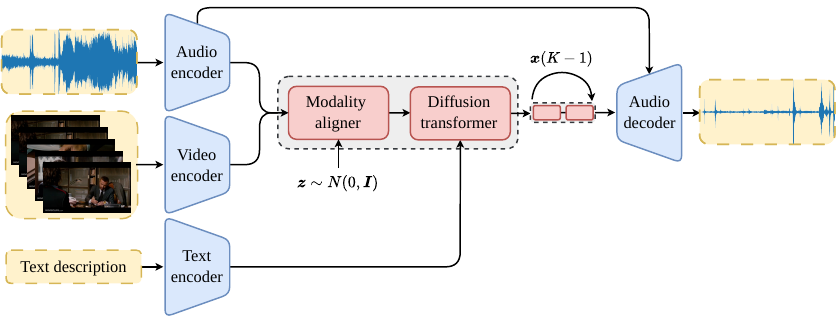}
    \caption{
        \textbf{Overview of \ac{\modelName}.}
        Given a video with degraded and unbalanced audio, \ac{\modelName} enhances it according to the given textual description.
        First, all the modalities are encoded with specific encoders, and the visual and audio features are aligned and fused.
        The fused and textual features are then passed to the \ac{DiT}.
        The \ac{DiT} generates enhanced audio that closely follows the original content and user guidance.
        Finally, the generated audio is decoded into a waveform representation with the SkipDACVAE decoder.
    }
    \label{fig:ssr}
\end{figure*}

\ac{\modelName}, presented in \cref{fig:ssr}, is a multimodal generative approach for \ac{UGC} enhancement.
Given a video from an everyday environment, \ac{\modelName} enhances the audio according to the user's preferences.
All modalities are first encoded separately.
Then, audio and visual features are temporally aligned and fused.
The fused features are passed to a generative model, where text guides generation via cross-attention, and the output is decoded into a waveform.
\ac{\modelName} can rebalance the audio mix, remove unwanted sources, and generate new audio content to replace degraded audio.

\subsection{Neural Audio Codec -- SkipDACVAE}
We propose SkipDACVAE, adapted from the DACVAE \cite{polyak2024movie,kumar2023high} audio codec.
DACVAE \cite{polyak2024movie} replaces the \ac{RVQ} bottleneck of the original DAC \cite{kumar2023high} with a \ac{VAE} bottleneck.
This gives DACVAE a continuous latent space which is well suited for continuous flow matching \cite{lipman2022flow}.

SkipDACVAE adds skip connections from the encoder to the decoder.
This improves reconstruction quality, since the decoder can access the original audio features directly.
Each skip connection adds encoder features to decoder features at matching temporal resolution.
Before addition, a 1D convolutional layer with kernel size of 1 maps the encoder features to the decoder's channel dimension.
Each connection uses its own convolutional layer.
The number of connections equals the number of downsampling layers in the encoder.

We initialize SkipDACVAE with pretrained DACVAE weights and freeze all shared layers, leaving only the added 1D convolutional layers trainable.
These layers are zero-initialized, so SkipDACVAE behaves identically to DACVAE at the start of training.
During training, we obtain a degraded version of the input audio and pass it through SkipDACVAE to compute the \textit{skipped} values.
We then encode the original, undegraded audio and decode it using \textit{skipped} values computed from corresponding degraded version.
We compute the reconstruction loss between the decoded output and the original audio, and backpropagate it to update SkipDACVAE.
Using the degraded audio to compute \textit{skipped} values forces SkipDACVAE to transfer high-level information from encoder to decoder, rather than simply copying the original audio features.
We use the original training configuration of DAC \cite{kumar2023high}.

During training and inference of \ac{\modelName}, we use a pretrained SkipDACVAE.
We encode the input audio mixture, yielding a latent sequence $\vx_{mix} \in \R^{T \times C_{aud}}$ at 25 Hz, where $T$ is the number of audio frames and $C_{aud} = 128$ is the channel dimension.

\vspace{-8pt}
\subsection{Visual Encoder}
We utilize a pretrained \ac{PE} \cite{bolya2025perception} as the visual encoder.
\ac{PE}'s large scale contrastive training on vision-language pairs allows it to learn semantically rich representations for actions and scene context.
We encode the video per-frame, yielding a sequence of visual features $\vx_{vis} \in \R^{{C_{vis}} \times T_{vis}}$, where $T_{vis}$ is the number of visual frames and $C_{vis} = 1024$ is the channel dimension.

\subsection{Text Encoder}
\label{ssec:text_encoder}
Text descriptions are encoded with a pretrained T5-base encoder \cite{raffel2020exploring}.
Text feature sequence $\vx_{text} \in \R^{N \times C_{text}}$, where $N$ is the number of text tokens and $C_{text} = 768$ is the channel dimension, is obtained from the last hidden layer of the encoder.
Before feeding the text features to the cross-attention layers of the generative model, we project them to the model's channel dimension $C$ using a linear layer.

\subsection{Generative Conditional Flow Matching Model}
We propose a generative approach for audio remixing based on \ac{CFM} \cite{lipman2022flow} and recent audio source separation techniques \cite{shi2025samaudio}.
Our model learns a continuous vector field that transports a Gaussian prior sample $\vx_0 \sim N(0, \mI)$ to a target data sample $\vx_1$ over $t \in [0, 1]$, where $t$ is the timestep.
At each timestep, the model predicts a velocity field $\vv_{\theta}(t, \mC, \vx_t)$, where $\mC = \{\vx_{mix},\vx_{vis},\vx_{text}\}$ is the conditioning signal, $\vx_t$ is the flow state at time $t$, and  $\theta$ are the model parameters.
By integrating $\vv_{\theta}$ over time, we obtain the final prediction $\hat\vx_1$.

\PAR{Modality Aligner.}
The target data sample $\vx_1 \in \R^{T \times 2C_{aud}}$ stacks the target audio $\vx_{tgt}$ and the residual signal $\vx_{mix} - \vx_{tgt}$ along the channel dimension.
The Gaussian prior sample $\vx_0 \in \R^{T \times 2C_{aud}}$ stacks noise $\vz \sim N(0, \mI)$ and an all-zero tensor in the same way.
Following \cite{shi2025samaudio}, we condition the model on the audio mixture $\vx_{mix}$ for content-faithful generation.
At each flow step, we concatenate the flow state $\vx_t$ with the fixed mixture $\vx_{mix}$, then project to the model's channel dimension $C$ using a linear layer, yielding $\vh \in \R^{T \times C}$.

We also add visual information to $\vh$.
First, to temporally align the audio and visual features, we match the frame rate of $\vx_{vis}$ to 25 Hz using nearest-neighbor interpolation, yielding a visual feature sequence $\vx_{vis} \in \R^{C_{vis} \times T}$.
Then, visual information is projected and added via a gated summation: $\vh = \vh + \tanh(\sigma) \odot \mathrm{LN}(\mathrm{Conv1D}(\vx_{vis})^\top)$, where $\sigma$ is a learnable gating parameter, $\odot$ denotes elementwise (broadcasted) multiplication, $\mathrm{LN}$ is layer normalization, and $\mathrm{Conv1D}$ is a 1D convolutional layer with kernel size of 1, that projects the visual features to the model's channel dimension $C$.

\PAR{Diffusion Transformer.}
We represent the velocity field with a \ac{DiT} \cite{peebles2023scalable} and previously introduced \textit{Modality aligner}.
\ac{\modelName} adopts an architecture where each Transformer \cite{vaswani2017attention} block is modulated by the flow time embedding via scale-and-shift operations on normalization and residual layers.
To create the flow time embedding, a shared MLP maps $t$ to six modulation parameters (four scales, two biases), reused across blocks with layer-specific biases added for depth-dependent effects, reducing model size.

In practice, we sample $t$ as $t = k / (K-1)$, where $k \sim \mathcal{U}\{0, \dots, K-1\}$ and $K=4$.
As shown in previous works \cite{chadebec2025lbm,chadebec2025flash,luo2023latent, salimans2022progressive}, selecting a few timesteps during training may be beneficial during inference time in image-to-image generation tasks.
Also, a concurrent work by Malard et al. \cite{malard2026conditional} utilize a similar timestep sampling approach.
$K=4$ yielded the best overall results in our testing, and we use this value for all experiments.
During inference, we use Euler integration with the same amount of steps.

We initialize the model with pretrained weights of SAM-Audio \cite{shi2025samaudio}.
We argue that pretraining the model on a large-scale audio source separation task is beneficial for remixing and enhancement, as it allows the model to learn general audio representations and source separation capabilities.
In our experiments, training the query, key, value, and output projections across all \ac{DiT} blocks together with the \textit{Modality aligner} and the linear layer projecting the $\vx_{text}$ features (\cref{ssec:text_encoder}) yielded the best results.
We compared different selective unfreezing strategies, LoRA \cite{hu2022lora} and full finetuning.

%%% ---------------------------------

\section{Experiments}
\label{sec:experiments}
%%% ---- Section Content: Experiments ----
\subsection{Dataset -- DegradedMix}
We base our data on the MuddyMix dataset \cite{huang2025learning}, also used by the compared approaches \cite{huang2025learning,malard2026conditional}.
MuddyMix consists of 15,078/1,927/1,789 train/validation/test 10-second clips extracted from movies \cite{bain2020condensed}, each with a professionally mixed audio track.
Each track is separated into stems via a pretrained source separation model \cite{Watcharasupat2024RemasteringDA}, $s = \hat{s}_h + \hat{s}_m + \hat{s}_e + \hat{s}_r$, denoting speech, music, effects, and residual signals.
MuddyMix degrades these stems with random discrete gain changes, then remixes them into a degraded audio track.

To mimic real-world \ac{UGC} enhancement scenarios, we propose \emph{DegradedMix} which extends MuddyMix by adding background noise \cite{Wichern2019WHAM} and reverberation \cite{sox} to the audio mixes.
We train on DegradedMix and evaluate on both DegradedMix (\cref{tab:rev_noise_metrics,tab:rev_noise_mm_metrics}) and MuddyMix (\cref{tab:muddymix_metrics,tab:genai_metrics}) test sets, enabling comparison to prior work under standard and more challenging degradations.
Note that MuddyMix \cite{huang2025learning} does not support background noise removal, as all the stems in the degraded mix are also present in the ground-truth mix.

We also add audio-based captions to the DegradedMix dataset, which describe the original, non-degraded audio content.
One-to-two sentence audio captions are generated automatically with Qwen2-Audio \cite{Qwen2-Audio}.
This mimics the real-world scenario where the user provides a textual description of the desired enhancement.
We release the code to generate DegradedMix data.

\subsection{Compared Methods}
The proposed model is compared against two recent approaches, namely VisAH \cite{huang2025learning} and VisAH-FM \cite{malard2026conditional}.
VisAH is a discriminative approach that uses a Transformer-based \cite{vaswani2017attention} encoder-decoder model to predict a mask in spectral space which is applied to the input audio to generate the remixed audio.
VisAH-FM is a closed-source concurrent work that was published after the original VisAH which uses a conditional flow matching approach to generate the remixed audio, following a similar architecture as VisAH.

\subsection{Implementation and Training Details}
We train on 25 FPS video with 48 kHz audio, resampled to 44.1 kHz for evaluation to match prior work \cite{huang2025learning,malard2026conditional}.
Video frames are resized to $336 \times 336$, and \ac{\modelName} uses a 10-second context length, following \cite{huang2025learning,malard2026conditional}.
We generate degraded audio on-the-fly during training, applying gain changes to the stems and adding random background noise and reverberation.
The model is initialized from pretrained weights \cite{shi2025samaudio} and trained for approximately 67K steps with batch size 26 on two NVIDIA RTX PRO 6000 Blackwell GPUs, using AdamW \cite{loshchilov2019decoupled} ($\beta=[0.9,0.95]$, weight decay 0.1, learning rate $1e^{-4}$) with a cosine annealing schedule and 2K warmup steps.

\subsection{Evaluation Metrics}
The evaluation protocol proposed by Huang et al. \cite{huang2025learning} is largely based on comparing the automatic remix with the ground truth balanced mix.
These metrics include a) magnitude \cite{xu2021visually} and envelope \cite{liang2023av} distance between the mixes, b) time alignment using Wasserstein distance, and c) semantic alignment using \ac{KL} \cite{liu2024audioldm, vyas2023audiobox, koutini22passt} and \acf{IB} \cite{girdhar2023imagebind, viertola2025temporally}.
In particular, a) and b) are ill-suited for creative tasks like remixing, where multiple valid solutions can exist.
We instead adopt evaluation metrics from generative modeling, which better suit the setting.

Following evaluation practices in multimodal audio generation \cite{liu2024audioldm,cheng2025mmaudio}, we propose to use the following metrics to evaluate the remixing performance: a) \ac{FD} \cite{kilgour2018fr} and \acf{KL} \cite{liu2024audioldm, vyas2023audiobox} for distribution matching, c) \ac{IS} \cite{salimans2016improved} to measure audio quality, d) \acf{IB} \cite{girdhar2023imagebind} for semantic alignment, e) \ac{Sync} \cite{iashin2024synchformer}, and f) \ac{CLAP} \cite{wu2023large}.
For distribution matching scores, we utilize PANNs \cite{kong2020panns} and PaSST \cite{koutini22passt} pretrained audio classifiers to extract features from the generated remixes and the ground truth mixes.
For \ac{IS}, we use the same PANNs classifier.

\subsection{Results}
% \begin{table}[th]
%     \centering
%     \renewcommand{\tabcolsep}{4pt}
%     \caption{
%         Original metrics for DegradedMix test data.
%         We could not evaluate VisAH-FM \cite{malard2026conditional} as the model is closed-source.
%         *: retrained with the DegradedMix data for fair comparison.
%     }
%     \begin{tabular}{l l c c c c}
%         \toprule
%         Variant & & Env $\downarrow$ & Mag $\downarrow$ & Was $\downarrow$ & $\text{KL}_{\mathrm{PaSST}} \downarrow$\\
%         \midrule
%         \textit{Input} & & 6.79 & 23.13 & 1.95 & 25.65 \\
%         VisAH \cite{huang2025learning} & & 5.16 & 15.70 & 1.22 & 21.08 \\
%         \ac{\modelName} & & 3.65 & 10.87 & 0.90 & 11.60 \\
%         \bottomrule
%     \end{tabular}
%     \label{tab:abl_modelsize}
% \end{table}

\begin{table}[th]
    \centering
    \renewcommand{\tabcolsep}{1.6pt}
    \caption{
        Proposed metrics calculated for DegradedMix test data.
        We could not evaluate VisAH-FM \cite{malard2026conditional} as the model is closed-source.
        *: retrained with the DegradedMix data for fair comparison.
    }
    \vspace{-5pt}
    \begin{tabular}{l l c c l c c c c c c}
        \toprule

        & &
        \multicolumn{2}{c}{\textit{PANNs}} &
        &
        \multicolumn{2}{c}{\textit{PaSST}} &
        & & & \\

        \cmidrule{3-4}\cmidrule{6-7}

        \multirow{-2}{*}{Model} & &
        FD $\downarrow$ &
        KL $\downarrow$ &
        &
        FD $\downarrow$ &
        KL $\downarrow$ &
        \multirow{-2}{*}{IS $\uparrow$} &
        \multirow{-2}{*}{IB $\uparrow$} &
        \multirow{-2}{*}{Sync $\downarrow$} &
        \multirow{-2}{*}{CLAP $\uparrow$} \\
        \midrule

        \textit{Input} & & 3.55 & 0.35 & & 43.39 & 0.26 & \textbf{3.16} & 30.49 & 51.80 & 40.56 \\
        VisAH* \cite{huang2025learning} & & 2.97 & 0.31 & & \textbf{40.63} & 0.21 & 3.04 & 31.08 & 51.80 & 44.06 \\
        \ac{\modelName} & & \textbf{1.26} & \textbf{0.14} & & 41.30 & \textbf{0.12} & 3.13 & \textbf{32.44} & \textbf{48.67} & \textbf{45.06} \\

        \bottomrule
    \end{tabular}
    \label{tab:rev_noise_metrics}
\end{table}

% WandB ID: tufvq8c7

\begin{table}[th]
    \centering
    \renewcommand{\tabcolsep}{6.9pt}
    \caption{
        \ac{VGAH} \cite{huang2025learning} metrics calculated for DegradedMix test data.
        We could not evaluate VisAH-FM \cite{malard2026conditional} as the model is closed-source.
        *: retrained with the DegradedMix data for fair comparison.
    }
    \vspace{-5pt}
    \begin{tabular}{l l c c c c}
        \toprule
        Variant & & Env $\downarrow$ & Mag $\downarrow$ & Was $\downarrow$ & $\text{KL}_{\mathrm{PaSST}} \downarrow$\\
        \midrule
        \textit{Input} & & 6.79 & 23.13 & 1.95 & 25.65 \\
        VisAH* \cite{huang2025learning} & & 5.16 & 15.70 & 1.22 & 21.08 \\
        \ac{\modelName} & & \textbf{3.65} & \textbf{10.87} & \textbf{0.90} & \textbf{11.60} \\
        \bottomrule
    \end{tabular}
    \label{tab:rev_noise_mm_metrics}
\end{table}

% \begin{table}[th]
%     \centering
%     \renewcommand{\tabcolsep}{1.75pt}
%     \caption{
%         Proposed metrics calculated for DegradedMix test data.
%         We could not evaluate VisAH-FM \cite{malard2026conditional} as the model is closed-source.
%         *: retrained with the DegradedMix data for fair comparison.
%     }
%     \begin{tabular}{l l c c l c c c c c c}
%         \toprule

%         & &
%         \multicolumn{2}{c}{\textit{PANNs}} &
%         &
%         \multicolumn{2}{c}{\textit{PaSST}} &
%         & & & \\

%         \cmidrule{3-4}\cmidrule{6-7}

%         \multirow{-2}{*}{Model} & &
%         FD $\downarrow$ &
%         KL $\downarrow$ &
%         &
%         FD $\downarrow$ &
%         KL $\downarrow$ &
%         \multirow{-2}{*}{IS $\uparrow$} &
%         \multirow{-2}{*}{IB $\uparrow$} &
%         \multirow{-2}{*}{Sync $\downarrow$} &
%         \multirow{-2}{*}{CLAP $\uparrow$} \\
%         \midrule

%         \textit{Input} & & 3.55 & 0.35 & & 43.39 & 0.26 & \textbf{3.16} & 30.49 & 51.80 & 40.56 \\
%         VisAH* \cite{huang2025learning} & & 2.97 & 0.31 & & \textbf{40.63} & 0.21 & 3.04 & 31.08 & 51.80 & 39.95 \\
%         \ac{\modelName} & & \textbf{1.26} & \textbf{0.14} & & 41.30 & \textbf{0.12} & 3.13 & \textbf{32.44} & \textbf{48.67} & \textbf{45.06} \\

%         \bottomrule
%     \end{tabular}
%     \label{tab:rev_noise_metrics}
% \end{table}

% WandB ID: tufvq8c7

\PAR{Audio Enhancement.}
\Cref{tab:rev_noise_metrics} reports model performance on DegradedMix, which includes gain imbalance, background noise, and reverberation as degradations, using the proposed metrics.
For a fair comparison, we retrain VisAH \cite{huang2025learning} on DegradedMix, and also report metrics for the unprocessed degraded \textit{Input}.
We could not evaluate VisAH-FM \cite{malard2026conditional}, as its code is not publicly available.
Our model outperforms VisAH across all metrics except PaSST \cite{koutini22passt} \ac{FD}, highlighting its effectiveness in enhancing degraded audio.
The high \ac{CLAP} and \ac{IB} scores further show the model's ability to follow conditional guidance, which is crucial for user-guided audio enhancement.

\Cref{tab:rev_noise_mm_metrics} reports the model performance on DegradedMix using the \ac{VGAH} metrics \cite{huang2025learning}.
Our model clearly outperforms VisAH.

\begin{table}[th]
    \centering
    \renewcommand{\tabcolsep}{1.75pt}
    \caption{
        Proposed metrics calculated for MuddyMix \cite{huang2025learning} test data.
        We could not evaluate VisAH-FM \cite{malard2026conditional} as the model is closed-source and the authors do not provide samples for the MuddyMix test set.
        % $^\dagger$: VisAH variant conditioned on video frame captions.
    }
    \vspace{-5pt}
    \begin{tabular}{l l c c l c c c c c c}
        \toprule

        & &
        \multicolumn{2}{c}{\textit{PANNs}} &
        &
        \multicolumn{2}{c}{\textit{PaSST}} &
        & & & \\

        \cmidrule{3-4}\cmidrule{6-7}

        \multirow{-2}{*}{Model} & &
        FD $\downarrow$ &
        KL $\downarrow$ &
        &
        FD $\downarrow$ &
        KL $\downarrow$ &
        \multirow{-2}{*}{IS $\uparrow$} &
        \multirow{-2}{*}{IB $\uparrow$} &
        \multirow{-2}{*}{Sync $\downarrow$} &
        \multirow{-2}{*}{CLAP $\uparrow$} \\
        \midrule

        \textit{Input} & & 2.30 & 0.29 & & 31.86 & 0.21 & 3.11 & 31.19 & 51.19 & 38.30 \\
        VisAH \cite{huang2025learning} & & 1.26 & 0.18 & & \textbf{18.06} & 0.11 & 3.06 & 31.81 & 49.71 & 39.56 \\
        % VisAH$^\dagger$ \cite{huang2025learning} & & 1.26 & 0.18 & & \textbf{17.90} & 0.11 & 3.05 & 31.91 & 50.07 & 38.74 \\
        \ac{\modelName} & & \textbf{1.04} & \textbf{0.13} & & 35.78 & \textbf{0.10} & \textbf{3.13} & \textbf{32.74} & \textbf{47.69} & \textbf{46.78} \\

        \bottomrule
    \end{tabular}
    \label{tab:genai_metrics}
\end{table}

% WandB ID: tufvq8c7

\begin{table}[th]
    \centering
    \renewcommand{\tabcolsep}{6.9pt}
    \caption{
        \ac{VGAH} \cite{huang2025learning} metrics calculated for MuddyMix \cite{huang2025learning} test data.
        % $^\dagger$: VisAH \cite{huang2025learning} variant conditioned on video frame captions.
    }
    \vspace{-5pt}
    \begin{tabular}{l l c c c c c}
        \toprule
        Model & & Env $\downarrow$ & Mag $\downarrow$ & Was $\downarrow$ & $\text{KL}_{\mathrm{PaSST}} \downarrow$ \\ % & IB Score $\uparrow$\\
        \midrule
        \textit{Input} & & 6.29 & 22.69 & 1.96 & 20.74 \\ % & 28.14 \\ % from VisAH-FM paper
        VisAH \cite{huang2025learning} & & 3.38 & 9.99 & 0.84 & 11.37 \\ % & 28.84 \\ % from VisAH-FM paper
        % VisAH$^\dagger$ \cite{huang2025learning} & & 3.44 & 10.22 & 0.88 & 11.71 \\ % & 28.92 \\ % from VisAH-FM paper
        % VisAH \cite{huang2025learning} & & 3.67 & 10.20 & 0.81 & 11.18 & 25.00 \\ % for the actual generated results
        % VisAH$^\dagger$ \cite{huang2025learning} & & 3.66 & 10.04 & 0.80 & 11.01 &  25.07 \\ % for the actual generated results
        VisAH-FM \cite{malard2026conditional} & & \textbf{2.74} & \textbf{8.28} & \textbf{0.63} & 9.70 \\ % & 29.12 \\
        \ac{\modelName} & & 3.46 & 9.86 & 0.88 & \textbf{9.61} \\ % & \textbf{32.74} \\ % this is the same as in the genai_metrics table
        \bottomrule
    \end{tabular}
    \label{tab:muddymix_metrics}
\end{table}

% WandB ID: tufvq8c7
% Ours frame-wise IB-Score: 0.2586921456382429

% VisAH frame-wise IB-Score: 0.24998979829824408
% VisAH-dagger frame-wise IB-Score: 0.25071146232562735

% For some reason I can not match the frame-wise IB Score reported in the VisAH-FM paper for the VisAH-model...
% Contacted the authors about it. Other metrics align pretty close, so IB Score should not be that much off.

\PAR{Audio Remixing.}
\Cref{tab:genai_metrics} reports the model performance on MuddyMix, which includes only gain imbalance as a degradation, using the proposed metrics.
We could not evaluate VisAH-FM \cite{malard2026conditional} as its code or samples are not available.
\ac{\modelName} exceeds all the compared methods across all the metrics except PaSST \cite{koutini22passt} \ac{FD}.
However, we clearly achieve the best balance between all the metrics.

\Cref{tab:muddymix_metrics} reports model performance on MuddyMix \cite{huang2025learning} using \ac{VGAH} metrics.
When compared against a single ground-truth solution (Env, Mag, Was), \ac{\modelName} falls short of the concurrent work VisAH-FM \cite{malard2026conditional}, though we achieve the best $\text{KL}_{\mathrm{PaSST}}$ score.
We argue this reflects a limitation of the metric rather than the model: in a creative task like remixing, where multiple valid solutions can exist, comparing generated results to a single ground truth is not ideal.
This is further highlighted by the subjective evaluation in \cref{fig:preference_rates}, where our approach (\ac{\modelName}-L) is preferred over VisAH with a large margin, despite scoring only marginally better or on par under \ac{VGAH} metrics.

\begin{table}[h]
    \centering
    \renewcommand{\tabcolsep}{4pt}
    \caption{
        Pairwise preference test results for \ac{UGC} enhancement. Preference rate is the proportion of trials in which the first condition was preferred over the second, with 95\% confidence intervals.
        $p$-values are from a two-sided binomial test against chance (50\%).
    }
    \vspace{-5pt}
    \label{fig:preference_rates}
    \begin{tabular}{lccccc}
        \toprule
        Comparison (A vs. B) & Wins (A--B) & Pref.\ rate & $p$-value \\
        \midrule
        \ac{\modelName}-L vs.\ \ac{\modelName}-S              & 68--22 & 75.6\% & $< 0.001$ \\
        \ac{\modelName}-L vs.\ Original audio                 & 72--18 & 80.0\% & $< 0.001$ \\
        \ac{\modelName}-L vs.\ VisAH \cite{huang2025learning} & 80--10 & 88.9\% & $< 0.001$ \\
        \ac{\modelName}-S vs.\ Original audio                 & 47--43 & 52.2\% & $0.752$   \\
        \ac{\modelName}-S vs.\ VisAH \cite{huang2025learning} & 57--33 & 63.3\% & $0.015$   \\
        Original audio vs.\ VisAH \cite{huang2025learning}    & 62--28 & 68.9\% & $< 0.001$ \\
        \bottomrule
    \end{tabular}
\end{table}

\PAR{Subjective Analysis.}
We conduct a subjective evaluation on \ac{UGC} audio enhancement.
We select 10 \ac{UGC} videos (presented at the project page) and generate enhanced audio using the small- and large-variants of \ac{\modelName} (\ac{\modelName}-S and \ac{\modelName}-L), and VisAH \cite{huang2025learning}.
10 participants select the preferred audio for a video between two samples, presented in random order.
Preference rate tells how many times on average a particular audio was preferred over other options.
\Cref{fig:preference_rates} shows the final rates, where \ac{\modelName}-L is preferred in 81\% of the cases, demonstrating its effectiveness.
\ac{\modelName}-S is preferred in 47\% and the original audio is preferred in 46\% of the cases.
VisAH's limited \ac{UGC} enhancement and remixing capabilities result in a lower preference rate of 26\%.

\subsection{Ablations}

% \begin{table}[th]
%     \centering
%     \renewcommand{\tabcolsep}{4pt}
%     \caption{Ablation on different model variants.}
%     \begin{tabular}{l l c c c c}
%         \toprule
%         Variant & & Env $\downarrow$ & Mag $\downarrow$ & Was $\downarrow$ & $\text{KL}_{\mathrm{PaSST}} \downarrow$\\
%         \midrule
%         Small & & 3.45 & 10.28 & 0.83 & 10.62 \\
%         Base & & 3.40 & 10.03 & 0.85 & 10.11 \\
%         \rowcolor{yellow!20}
%         Large & & 3.46 & 9.86 & 0.88 & 9.61 \\
%         \bottomrule
%     \end{tabular}
%     \label{tab:abl_modelsize}
% \end{table}

\begin{table}[H]
    \centering
    \renewcommand{\tabcolsep}{1.75pt}
    \caption{
        Ablation on different model variants.
        We use MuddyMix \cite{huang2025learning} data.
        Preferred configuration is highlighted in yellow.
    }
    \vspace{-5pt}
    \begin{tabular}{l l c c l c c c c c c}
        \toprule

        & &
        \multicolumn{2}{c}{\textit{PANNs}} &
        &
        \multicolumn{2}{c}{\textit{PaSST}} &
        & & & \\

        \cmidrule{3-4}\cmidrule{6-7}

        \multirow{-2}{*}{Model} & &
        FD $\downarrow$ &
        KL $\downarrow$ &
        &
        FD $\downarrow$ &
        KL $\downarrow$ &
        \multirow{-2}{*}{IS $\uparrow$} &
        \multirow{-2}{*}{IB $\uparrow$} &
        \multirow{-2}{*}{Sync $\downarrow$} &
        \multirow{-2}{*}{CLAP $\uparrow$} \\
        \midrule

        Small & & 1.12 & 0.14 & & 37.81 & 0.11 & 3.13 & 32.76 & 47.82 & 47.02 \\
        Base & & 1.06 & 0.13 & & 37.28 & 0.10 & 3.12 & 32.64 & 48.84 & 47.10 \\
        \rowcolor{yellow!20}
        Large & & 1.04 & 0.13 & & 35.78 & 0.10 & 3.13 & 32.74 & 47.69 & 46.78 \\

        \bottomrule
    \end{tabular}
    \label{tab:abl_modelsize}
\end{table}

% WandB IDs
% Large: tufvq8c7
% Base: f1ow7sf2
% Small: yxseqs11

\PAR{Model Size.}
\Cref{tab:abl_modelsize} presents the performance of different model sizes.
Small-variant has 500M parameters, base-variant has 1B parameters, and large-variant has 3B parameters.
The approach is scalable and we observed that the large-variant yields the best performance across the majority of the metrics.
We conduct all the experiments using the large-variant of the model if not otherwise specified.

% \begin{table}[th]
%     \centering
%     \renewcommand{\tabcolsep}{4pt}
%     \caption{Ablation on different model variants.}
%     \begin{tabular}{l l c c c c}
%         \toprule
%         Variant & & Env $\downarrow$ & Mag $\downarrow$ & Was $\downarrow$ & $\text{KL}_{\mathrm{PaSST}} \downarrow$\\
%         \midrule
%         DACVAE & & 4.21 & 15.68 & 0.90 & 28.15 \\
%         \rowcolor{yellow!20}
%         SkipDACVAE & & 3.46 & 9.86 & 0.88 & 9.61 \\
%         \bottomrule
%     \end{tabular}
%     \label{tab:abl_modelsize}
% \end{table}

\begin{table}[th]
    \centering
    \renewcommand{\tabcolsep}{0.95pt}
    \caption{
        Effect of using SkipDACVAE.
        We use MuddyMix \cite{huang2025learning} data.
        Preferred configuration is highlighted in yellow.
    }
    \vspace{-5pt}
    \begin{tabular}{l l c c l c c c c c c}
        \toprule

        & &
        \multicolumn{2}{c}{\textit{PANNs}} &
        &
        \multicolumn{2}{c}{\textit{PaSST}} &
        & & & \\

        \cmidrule{3-4}\cmidrule{6-7}

        \multirow{-2}{*}{Variant} & &
        FD $\downarrow$ &
        KL $\downarrow$ &
        &
        FD $\downarrow$ &
        KL $\downarrow$ &
        \multirow{-2}{*}{IS $\uparrow$} &
        \multirow{-2}{*}{IB $\uparrow$} &
        \multirow{-2}{*}{Sync $\downarrow$} &
        \multirow{-2}{*}{CLAP $\uparrow$} \\
        \midrule

        DACVAE & & 3.38 & 0.21 & & 143.55 & 0.28 & 3.03 & 32.50 & 48.35 & 48.13 \\
        \rowcolor{yellow!20}
        SkipDACVAE & & 1.04 & 0.13 & & 35.78 & 0.10 & 3.13 & 32.74 & 47.69 & 46.78 \\

        \bottomrule
    \end{tabular}
    \label{tab:abl_skipdacvae}
\end{table}

% WandB ID: tufvq8c7

\PAR{SkipDACVAE.}
As shown in \cref{tab:abl_skipdacvae}, adding skip connections boosts the performance significantly.
We conduct the experiments using the MuddyMix \cite{huang2025learning} data.
The added representation capacity of the neural audio codec allows for better reconstruction of the audio which is reflected in the improved performance across all metrics.
Note that the model is not retrained for the SkipDACVAE as these two neural audio codecs share the same latent space.

% \begin{table}[th]
%     \centering
%     \renewcommand{\tabcolsep}{4pt}
%     \caption{Ablation on different caption types.}
%     \begin{tabular}{l l c c c c}
%         \toprule
%         Type & & Env $\downarrow$ & Mag $\downarrow$ & Was $\downarrow$ & $\text{KL}_{\mathrm{PaSST}} \downarrow$\\
%         \midrule
%         Original & & 3.92 & 11.16 & 1.03 & 11.70 \\
%         Summarized & & 3.84 & 11.08 & 1.00 & 11.63 \\
%         \rowcolor{yellow!20}
%         Audio-based & & 3.77 & 10.94 & 0.98 & 11.52 \\
%         \bottomrule
%     \end{tabular}
%     \label{tab:abl_modelsize}
% \end{table}

\begin{table}[!h]
    \centering
    \renewcommand{\tabcolsep}{1.15pt}
    \caption{
        Ablation on different caption types.
        We use small-variant and MuddyMix \cite{huang2025learning}.
        Preferred configuration is highlighted in yellow.
    }
    \vspace{-5pt}
    \begin{tabular}{l l c c l c c c c c c}
        \toprule

        & &
        \multicolumn{2}{c}{\textit{PANNs}} &
        &
        \multicolumn{2}{c}{\textit{PaSST}} &
        & & & \\

        \cmidrule{3-4}\cmidrule{6-7}

        \multirow{-2}{*}{Type} & &
        FD $\downarrow$ &
        KL $\downarrow$ &
        &
        FD $\downarrow$ &
        KL $\downarrow$ &
        \multirow{-2}{*}{IS $\uparrow$} &
        \multirow{-2}{*}{IB $\uparrow$} &
        \multirow{-2}{*}{Sync $\downarrow$} &
        \multirow{-2}{*}{CLAP $\uparrow$} \\
        \midrule

        Original & & 1.24 & 0.16 & & 39.15 & 0.12 & 3.13 & 32.58 & 49.97 & 44.82 \\
        Summarized & & 1.22 & 0.16 & & 38.47 & 0.12 & 3.12 & 32.66 & 49.11 & 44.78 \\
        \rowcolor{yellow!20}
        % Audio-based & & 1.20 & 0.16 & & 37.80 & 0.12 & 3.13 & 32.64 & 49.56 & 46.99 \\
        Audio-based & & 1.12 & 0.14 & & 37.81 & 0.11 & 3.13 & 32.76 & 47.82 & 47.02 \\

        \bottomrule
    \end{tabular}
    \label{tab:abl_captions}
\end{table}

% WandB IDs
% Original: 8vrpjd0t
% Summarized: 9khrfiuo
% Audio: 85svkju3

\PAR{Captions.}
We ablate on different styles of textual descriptions in \cref{tab:abl_captions} and use the small-variant of the model with MuddyMix \cite{huang2025learning}.
MuddyMix dataset provides captions for each video frame, extracted with 1 FPS.
These \textit{Original} captions are long and contain unrelated information for audio enhancement.
\textit{Summarized} captions are generated by summarizing the original ones using Claude Sonnet 5 \cite{anthropic2026claudesonnet5}.
Finally, \textit{Audio-based} captions are generated by describing the original, non-degraded, audio content of the video using Qwen2-Audio \cite{Qwen2-Audio}.
We use audio-based captions, as they mimic the real-world scenario where users provide a description of the desired enhancement.

%%% --------------------------------------

\section{Conclusion}
\label{sec:conclusion}
%%% ---- Section Content: Conclusion ----
\ac{\modelName} shows that a single multimodal generative model can enhance \ac{UGC} audio.
It outperforms VisAH in controllability and quality under both generative and reconstruction-based metrics, though results are more mixed under single-ground-truth metrics, underscoring the need for evaluation that embraces remixing's one-to-many nature.
We hope DegradedMix and our evaluation framework support this shift.

%%% -------------------------------------

% \section*{Acknowledgments}
% The authors thank collaborators and reviewers for helpful feedback.
% This work was supported in part by institutional research resources.

% BiBTeX files (here: bibliography). The IEEEbib.bst bibliography
\bibliographystyle{IEEEbib}
\bibliography{bibliography}

\end{document}